\documentclass[
aip, prl,
amsmath,amssymb,
reprint,
twocolumn
]{revtex4-2}

\usepackage{lineno}
\usepackage{wrapfig}
\usepackage{xcolor}
\usepackage{caption}
\usepackage{subcaption}
\usepackage{graphicx}
\usepackage{colordvi}
\usepackage{hyperref}
\usepackage{cleveref}
\Crefformat{figure}{Fig.~#2#1#3}
\Crefmultiformat{figure}{Figs.~#2#1#3}{ and~#2#1#3}{, #2#1#3}{ and~#2#1#3}
\Crefname{figure}{Fig.}{Figs.}

\newcommand{\captionfonts}{\small}
\makeatletter  % Allow the use of @ in command names
\long\def\@makecaption#1#2{%
  \vskip\abovecaptionskip
  \sbox\@tempboxa{{\captionfonts #1: #2}}%
  \ifdim \wd\@tempboxa >\hsize
    {\captionfonts #1: #2\par}
  \else
    \hbox to\hsize{\hfil\box\@tempboxa\hfil}%
  \fi
  \vskip\belowcaptionskip}
\makeatother   % Cancel the effect of \makeatletter

\begin{document}

%%\begin{frontmatter}

%\title[inclusiveAI]{inclusiveAI: A machine learning model for hydrogen and deuterium structure function over a large $Q^2$ and Bjorken $x$ range}
\title[inclusiveAI]{inclAI: A machine learning representation of the $F_2$ structure function over all charted $Q^2$ and $x$ range}

\author{S.~Brown}
\affiliation{Virginia Polytechnic Institute and State University, Blacksburg, VA 24061}
\author{G.~Niculescu}
\affiliation{James Madison University, Harrisonburg, VA 22807}
\author{I.~Niculescu}
\affiliation{James Madison University, Harrisonburg, VA 22807}
\date{ \today}

\begin{abstract}
Structure function data provide insight into the nucleon quark distribution. They are relatively straightforward to extract from the world's vast, and growing, amount of inclusive lepto-production data. In turn, structure functions can be used to model the physical processes needed for planning and optimizing future experiments. In this paper a machine learning algorithm capable of predicting, using a unique set of parameters, the $F_2$ structure function, for four-momentum transfer $0.055 \leq Q^2 \leq 800.0$~GeV$^2$ and for Bjorken $x$ from $2.8 \times 10^{-5}$ to the pion threshold is presented.  The model was trained and reproduces  the hydrogen and the deuterium data at the 7~\% level, comparable with the average uncertainty of the experimental data. Extending the model to other nuclei or expanding the kinematic range are straightforward. The model is at least ten times faster than existing structure functions parameterizations, making it an ideal candidate for event generators and systematic studies.

\end{abstract}


\maketitle
%%\end{frontmatter}

%\linenumbers

%%%%%%%%%%%%%
\section{Introduction}

Inclusive electron scattering experiments have been used for more than fifty years to gain insight into the structure of subatomic particles. As far back as the 1960s this type of experiments, carried out at the Stanford Linear Accelerator (SLAC), provided the experimental evidence for the existence of quarks\ \cite{slac_paper_dis1, slac_paper_dis2}.

Starting with the experimental cross--section for inclusive scattering one can define and extract the so--called structure function(s), which parameterize the spatial extent of the target. Using the wealth of data accumulated, several structure function  models have been developed. Some of these models are predominantly phenomenological\ \cite{eric_model1, eric_model2, nmc_model} while others build the structure function starting from the underlying parton distribution functions (PDFs). ``Hybrid'' approaches that combine the different approaches traditionally used in the deep inelastic and resonance regimes are also available\ \cite{kulagin}. For a recent review of these see\ \cite{pdg} and references therein.

The current work uses machine learning (ML) to develop a structure function model, named ``inclusiveAI''. The model aims to provide accurate $F_2$ predictions over as large a Bjorken $x$ and four momentum transfer, $Q^2$, kinematic region as possible. This includes both deep inelastic scattering (DIS) data as well as resonance region data. The model is built {\it ab initio} to handle both hydrogen and deuterium data and it can be easily extended to heavier nuclei. The model does not make assumptions, implicit or explicit, about the data, and a unique set of parameters is used to predict the structure function regardless of the target or the kinematic regime. While this model does not directly provide PDFs, it is fast and reasonably accurate, making it attractive for use in applications where very large number of predictions are needed (event generators, acceptance simulation, radiative correction estimation).

This paper is structured as follows. In Section 2 we briefly review the basics of inclusive electron scattering and main approaches in structure function modeling. Section 3 describes the input data set while Section 4 introduces the machine learning approach used in this study. Section 5 presents the structure function results including the associated uncertainty studies that were undertaken. The last section presents our conclusions.

%%%%%%%%%%%%%%%%%%%%
\section{\label{sec:kinematics} Inclusive Electron Scattering and Structure Function Models.}

%\section{Kinematics and Models}

The data used in this study come primarily from fixed target charged lepton--nucleon scattering experiments: a lepton of energy $E$ scatters from a stationary nucleon and is detected at an angle $\vartheta$ with an energy $E'$, while the final hadronic state is not detected. In the one--photon exchange approximation the lepton--nucleon scattering process is mediated by the exchange of a virtual photon and can be represented by the Feynman diagram shown in Fig.\ \ref{fig:eNprocess}, where $l, l'$ are the incident and scattered leptons respectively, $N$ is the target nucleon (of mass $M$), and $X$ represents the recoiling system. Some of the kinematic variables used to describe the inclusive lepton-nucleon scattering process are: the four momentum transferred from the lepton to the target nucleon, $Q^2$, the fraction of the nucleon's momentum carried by the struck quark, $x$, the energy lost by the lepton, $\nu= E - E^\prime$, and  the invariant mass squared of the hadronic final state, $W^2$:

\begin{figure}
\begin{center}
\includegraphics[width=76mm]{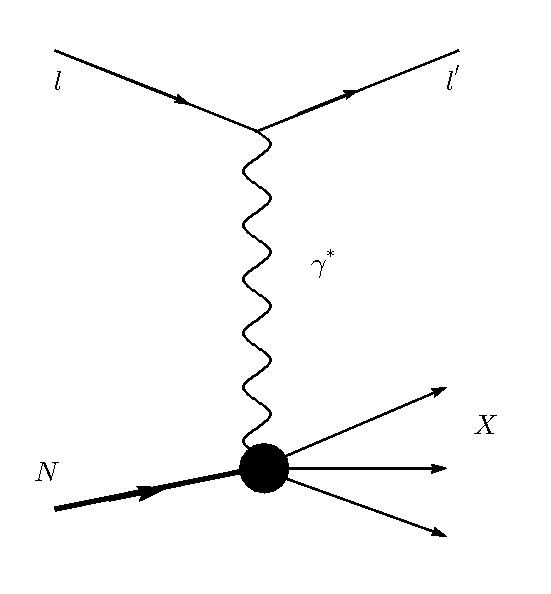}
\end{center}
  \caption{Feynman diagram for inclusive lepton--nucleon scattering in one--photon--exchange approximation.}
\label{fig:eNprocess}
\end{figure}

\begin{equation}
Q^2 = 4 E E^\prime \sin^2 \vartheta/2
\end{equation}

\begin{equation}
x  = \frac{Q^2}{2 M \nu}
\end{equation}

\begin{equation}
W^2 = M^2 + 2 M \nu - Q^2
\end{equation}

Using this approximation the differential cross section can be expressed in terms of structure functions $F_1$ and $F_2$, which parameterize the spatial extent of the charge distribution of partons inside the nucleon:

\begin{equation}
\begin{split}
\frac{d^2 \sigma}{d \Omega d E^\prime} = \sigma_{Mott} \Big( \frac{2}{M} F_1(x, Q^2) \tan^2 \frac{\vartheta}{2} \\ +\frac{1}{\nu} F_2(x, Q^2)\Big),
\end{split}
\end{equation} with $\sigma_{Mott}$ the cross section for scattering off of a point--like particle.
$F_1$ and $F_2$ can be related to the cross section for absorbing either a transversely ($\sigma_T$) or a longitudinally polarized virtual photon ($\sigma_L$)\ \cite{close}.

In the quark--parton model one can write the structure functions as combinations of the underlying quark  and anti--quark distribution functions. Various groups have used this formalism to extract parton distribution functions (PDFs). A list and brief discussion of the most recent PDF parameterizations available, including Machine Learning approaches,  can be found in\ \cite{pdg}. These parameterizations focus on the deep inelastic scattering process (large $Q^2$ and $W^2$) and are not suitable in the resonance region.
%{\it should we add something along the lines: these groups do not provide an easy way to compute experimental observables, like structure function???}

Phenomenological models of the inclusive cross section in the resonance region have been developed, with the most recent ones by Christy and Bosted\ \cite{eric_model1, eric_model2}. These models describe the cross section as a resonant contribution overlayed on top of a non-resonant background. The structure functions can then be obtained from the differential cross section using the ratio of the longitudinal and transverse cross sections, $R= \sigma_L/\sigma_T$.  

Both approaches can be computationally intensive when calculating experimental observables such as cross--section or structure functions. This is due to the convolutions required for each and every prediction. For PDF--based models this convolution is carried out over the parton distributions themselves. For the phenomenological parameterizations convolutions over the Fermi distribution are needed when calculating structure functions for deuteron/heavier nuclei. This can significantly impact several important data analysis steps where a very large number of predictions are needed, such as radiative correction estimation and detector response function modeling (i.e. acceptance calculations).

%as follows:
%\begin{equation}
%F_2 = \frac {d^2\sigma}{d\Omega dE'} \frac{1+R}{1+\varepsilon R}\times {kinem}.
%\end{equation}        
%\noindent where $kinem$ is a kinematic factor. 

%%%%%%%%%%%%%%%%%
\section{\label{sec:input_data} Input Data Selection}

% 01/30/2021
To develop the machine learning model described in this work the input data was selected and curated as follows:
\begin{itemize}
\item The data must be published or available from public sources/databases. 
\item The data must provide either the $F_2$ structure function or the differential cross--section. For the datasets providing only the latter the R1998 parameterization\ \cite{r1998} was used to extract $F_2$, with a small increase in the uncertainty budget. 
\item For each data point the statistical and systematic uncertainties, including any overall normalization errors, when known, were added in quadrature to obtain the total uncertainty. This was subsequently used to estimate the precision of the model as described in Section\ \ref{sec:fitting}. 
\item Only data above the pion threshold was used. 
\item No additional $Q^2$ or $x$ cuts were imposed, resulting in the most extensive data set available. 
\item Even though this study is limited to hydrogen and deuterium data, an extension to heavier targets is easily achievable. 
\end{itemize}

\begin{figure}
\begin{center}
\includegraphics[width=86mm]{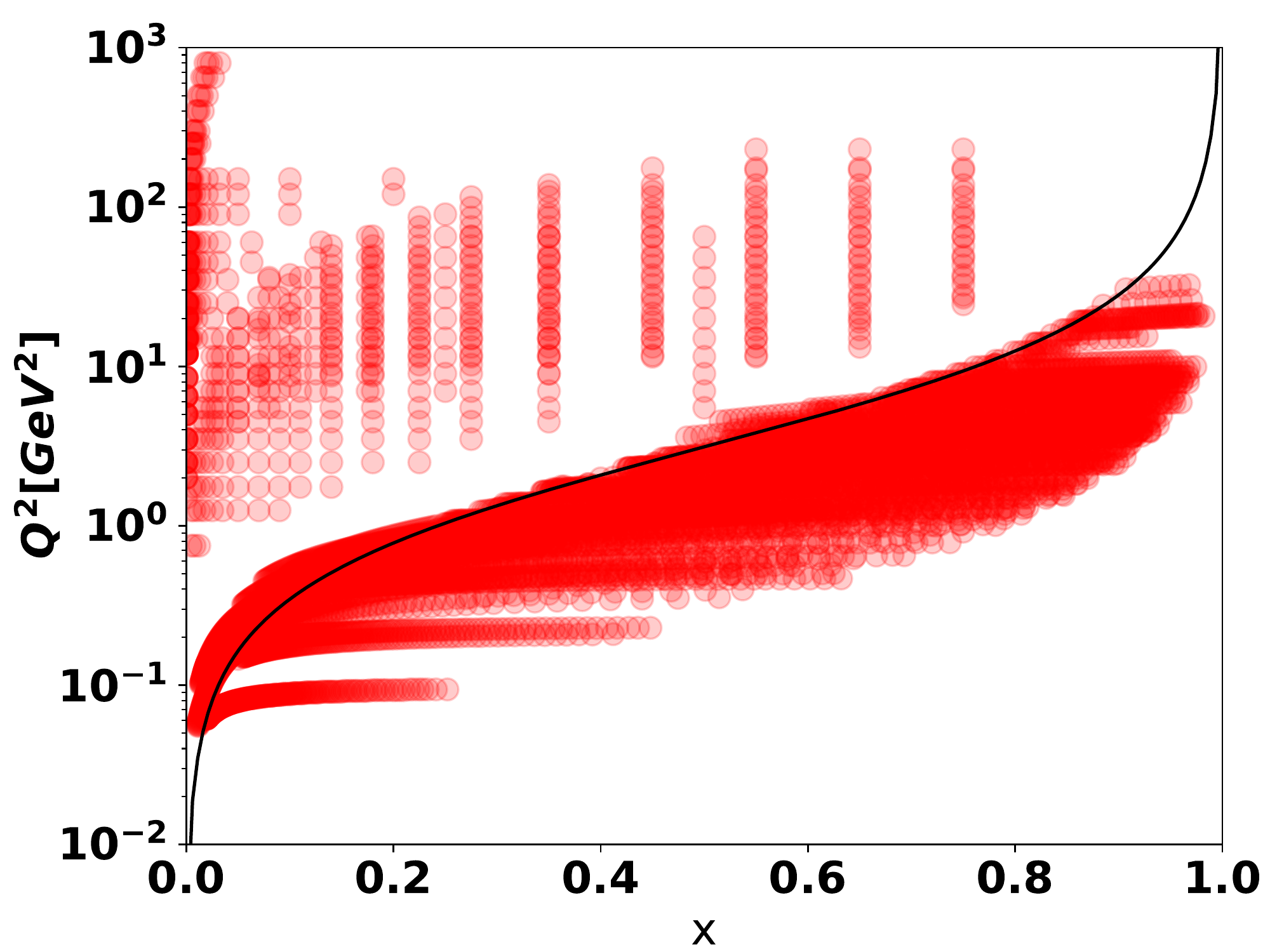}
\end{center}
  \caption{(Color online) $x$ and $Q^2$ coverage of the input hydrogen data set chosen for this study. The curve shown represents $W^2=4$~GeV$^2$; the nominal boundary between the resonance and the deep inelastic scattering regions.}
\label{fig:kinematics1}
\end{figure}

\begin{figure}
\begin{center}
\includegraphics[width=86mm]{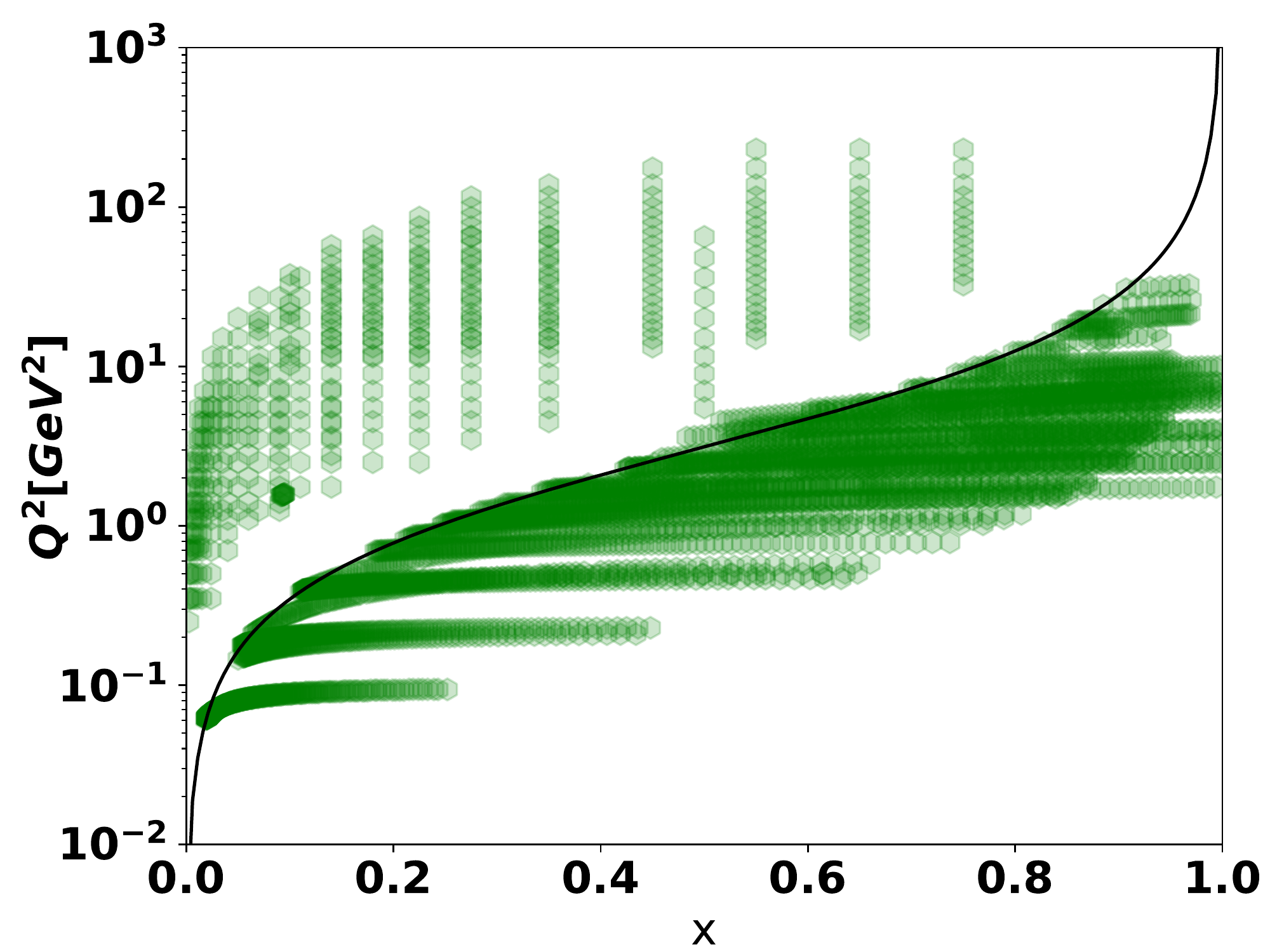}
\end{center}
  \caption{(Color online) $x$ and $Q^2$ coverage of the input deuterium data set chosen for this study. The curve shown represents $W^2=4$~GeV$^2$; the nominal boundary between the resonance and the deep inelastic scattering regions.}
\label{fig:kinematics2}
\end{figure}

The data set selected includes both electron--nucleon as well as muon--nucleon experiments, originating from several international laboratories:\\ SLAC\ \cite{slac1, slac2, slac3}, DESY\ \cite{desy}, CERN\ \cite{cern1, cern2, cern3}, and JLab\ \cite{jlab1, jlab2, jlab3, jlab4}. With such a large dataset, spanning several decades and laboratories, some ``tensions'' between datasets covering the same or adjacent phase space regions can be expected and have been documented\ \cite{slac1,tension2,tension3,tension4}. 
As it is difficult, if not impossible to carry out a full, {\it ab initio}, re-analysis of decades--old experiments in this study the input data was left ``as is'', not modified or renormalized {\it post hoc}. In other words, the input data was taken directly from the original publications stemming from the various experiments or from near--contemporaneous analyses thereof\ \cite{slac1}.  \Cref{fig:kinematics1,fig:kinematics2} show the coverage of the input data set chosen for this study for hydrogen and deuterium, respectively. The curve shown represents $W^2=4$~GeV$^2$, the nominal boundary between the resonance and the deep inelastic scattering regions.

A total of 11802 data points, 7612 on hydrogen and  4190 on deuterium targets, were used in this study. For both targets the bulk of the data is in the so--called ``resonance region'' ($W^2 < 4.0$~GeV$^2$), $\sim$80\% for hydrogen and $\sim$90\% for deuterium. In this study the absolute value of the deuteron $F_2$ structure function was used and not its relative value (i.e. the ``per nucleon'' $F_2$). The availability of a large number of data points in the region where the $F_2$ structure function exhibits substantial nonlinearity should help guide the training of the machine learning model.

%%%%%%%%%%%%%%%%%%%%%%%%%%%%%
\section{\label{sec:model} Machine Learning Model}

\subsection{\label{sec:rationale}
 Rationale}

As noted above, several approaches have been used, with varying degrees of success, to obtain fits of the structure function. All of these approaches assume a specific functional form (be it theory--inspired or phenomenological) for the fitted quantity. Subsequently, a minimization procedure is used to find the best values for the free parameters of this function. 

While in principle one can obtain the Hessian matrix associated with the fit and use error propagation to infer the uncertainties associated with quantities of interest that depend on the structure function (cross--section, moments), this procedure becomes impractical as the parameter space increases. Though a number of alternative methods circumventing the calculation of the full error matrix are available (truncated Newton, quasi--Newton methods\ \cite{nocedal}) one still runs the risk of potentially under- or over- estimating the uncertainty of the fit.

Furthermore, once  the choice of the functional form is made, one effectively has biased the algorithm against all other possible functions that share the same domain and codomain as the initial choice.  There are two substantial drawbacks associated with this choice, as described below.

First, lacking a clear theoretical guidance the functional form choice is somewhat arbitrary, and it requires constant refitting and/or redefinition (as in changing the functional form) as new data emerge. While for any data reduction problem the addition of a large body of new information does warrant the revision of one's model, the constant refitting of the model's parameters or the addition of new additional parameters casts doubts about a model's predictive power. 

Second, the functional form choice might limit the model's ability of making predictions over the whole domain of the fitted quantity, resulting in a model that performs extremely well but only on a subdomain of the available phase space. Some hybrid approaches attempt to address this issue by combining two/more models, each known to perform reasonably in its own portion of the structure function domain. The models are then combined using a designated ``merging phase space region'' where the prediction is simply a linear combination of the individual models. While this type of creative solution does work in practical implementations, the continuity of the prediction in the merging region suffers. 

Lastly, most of the theoretical insights are cast at the parton distribution level. For all cases where experimental observables, such as structure functions or cross--sections, 
need to be evaluated, cpu--intensive convolutions are required. This greatly increases the computational resources needed, especially for applications in which very large number of events are generated (Monte Carlo simulations, radiative corrections, etc.).

\subsection{\label{sec:ann} Artificial Neural Network architecture}

In recent years artificial intelligence/machine learning approaches have seen increased use in many fields, including substantial strides in nuclear and particle physics (pattern recognition, event reconstruction and topology, accelerator control and maintenance\ \cite{acc_control}). Significant strides have been made even in the specific area of structure function modeling\ \cite{nnpdf}.

The ML approach used here  attempts to address or circumvent the issues listed in\ \ref{sec:rationale} and provide fast, accurate structure function predictions for the structure function $F_2$. The model is completely data--driven, with no assumptions (or biases) of any underlying physics. The implementation is based on Artificial Neural Networks (ANN) and a back-propagation algorithm for the optimization of the network parameters.

The most important design constraints and features are presented below, grouped separately into physics choices and machine learning/implementation choices. As using ML is a relatively new addition to the computational capabilities of nuclear/particle physicists the latter set of choices will be more extensively detailed, highlighting the differences (and introducing some Data Analytics--specific vocabulary) between ML and traditional fitting procedures that readers might be familiar with.

Physics choices:
\begin{itemize}
\item the ML code shall provide $F_2$ predictions over as large  $Q^2$ (from $Q^2 < 1$ to $Q^2 \sim 1,000$ $GeV^2$) and $x$ (from very small ($\sim 10^{-5}$) to the pion threshold) as possible.
\item the ML input set shall incorporate all available charged lepton--nucleon data (DIS, resonance region, using electron as well as muon beams) that provides either the structure function $F_2$ itself or the differential cross--section (from which the structure function can be obtained). In the latter cases the generally accepted/used R1998\ \cite{r1998} function was used for the ratio of  the longitudinal and transverse cross--sections. 
\item the ML shall consider both hydrogen and deuterium data, with no bias or explicit provisions for either. Furthermore, the model shall provide a transparent way of generalizing the approach to other nuclei (see ML implementation below).
\item the total uncertainty associated with the input data shall be used in assessing the error associated with the ML predictions. No attempt to second--guess or re--scale the original, published, data shall be implemented. 
\end{itemize}

ML implementation choices:
\begin{itemize}
\item the ML model shall consist of an assembly of ANN with varying topologies. Simple majority voting (average) shall be used as the final ML prediction. Alternatively, one can pick a single topology as ``the'' model and use the spread in the $F_2$ predictions of the remaining architectures as a measure of the uncertainty.
\item the ML model shall be implemented using ``industry standard'' tools and libraries and should be able to complete its training using modest computation means.
\item the ML model shall run in a consistent manner and shall have a way of assessing the quality of its predictions.
\item the ML model uncertainties shall be commensurable/better than the total uncertainty associated with the input data points themselves.
\item the ML model shall try to minimize the mean square error (MSE) as it is an unbiased statistic.
\end{itemize}

From the ML/data analytics standpoint, $F_2$ prediction is a supervised learning exercise where  one seeks to infer the best possible values for the parameters of one's learning function given a set of ``labels'' (i.e. the observable(s) to be fitted, in this case $F_2$) and their corresponding ``features'' (i.e. the variables on which the observable depends). As the labels are known in advance for each set of inputs the ML method is deemed ``supervised''. Furthermore, given that the structure function $F_2$ takes real values, the ML algorithm is a regression rather than a classification.

Given that the behavior of the structure function, even considering the resonance region, is reasonably smooth and continuous, and given the time and hardware constraints of the project, a relatively simple ML was chosen, namely a fully connected ANN with one input layer, one output layer, and a number of hidden layers. While the number of neurons in the input and output layers are determined by the number of features and, respectively, by the number of labels (one), the number of hidden layers was varied: networks with up to ten hidden layers were tested. The number of neurons per layer was varied as well. Shallow networks (one or two hidden layers) required a relatively large number of neurons to achieve any significant performance. For one hidden layer network topologies with up to 1000 neurons/layer  were tested while for two hidden layer networks widths of up to $70 \times 70$ neurons were used. For networks deeper than two hidden layers topologies with 4 to 15 neurons/layer for layers two and above were used. As recommended in the machine learning literature\ \cite{geron} the number of neurons in the first layer was kept larger, 20 to 50 neurons. The number of neurons in layers two and above was kept the same for all layers. This substantially reduced the hyperparameter tuning task (less parameters that required study and optimization) without loss of model precision.

\begin{figure}
\begin{center}
\includegraphics[width=86mm]{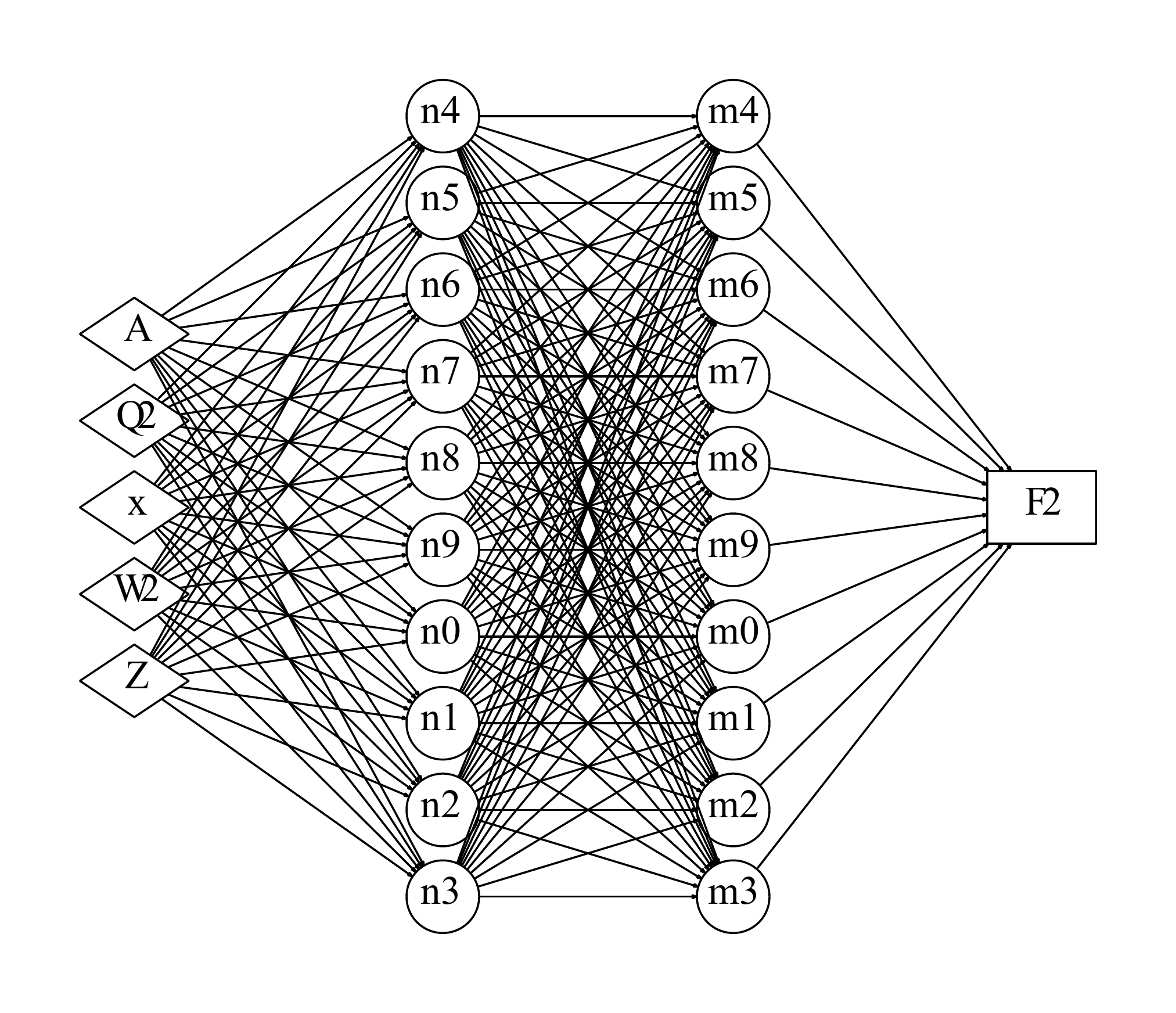}
\end{center}
  \caption{Sample topology for the Artificial Neural Networks used in this work.}
\label{fig:ann}
\end{figure}

Thus each network's topology can be summarized as 
\begin{equation}
(n_i, n_1, n_2, ..., n_m, n_o)
\end{equation}
where $n_i$ is the size of the input layer (i.e. the number of features used in the model), $n_o$ is the size of the output layer (the number of labels, which for this project was one, $F_2$), and $n_j$ are the sizes of the $m$ hidden layers. %For this study $1 \leq m \leq 10$, $n_1 \in [20,50]$, $n_{2 ... m} \in [4, ..., 15]$ for networks deeper than two hidden layers.
A graphical representation of one of the ANNs used in this work is shown in Fig.\ \ref{fig:ann}.

Theoretically, for a given target, $F_2$ is only a function of $x$ and $Q^2$. However, as the data set includes many resonance region entries, where several peaks are prominent (especially at low $Q^2$), $W^2$ was also added as a feature of the model. $W^2$ is, of course, not independent of $x$ and $Q^2$ and in traditional fitting approaches one seeks, as much as possible, independent input variables. ML models, however, often benefit from such ``feature engineering'' (i.e. creating new features based on existing ones) as one can model non-linear behavior using less neurons in the hidden layers than in the case when the network will only have the absolute minimum set of features. Lastly, as the model is to handle both hydrogen and deuterium data, the atomic ($Z$) and mass ($A$) numbers of the target were introduced as features. This brings the total number of features to five and also provides a straightforward path of extending the model to heavier nuclei if desired. %The addition of extra features (powers, positive or negative of $x$, $Q^2$, and $W^2$) was studied but 

Two types of activation functions were used in these networks: a sigmoid\ \cite{sigmoid_ref} for the output layer and a ReLu activation function, $f(x) = max(0,x)$\ \cite{relu_ref},  for the hidden layers. The input layer simply copies its input to its output.  The model outlined above was implemented in Python, using the Keras package\ \cite{keras} with a Tensorflow backend\ \cite{tf2}.

%%%%%%%%%%%%%%%%%%%%%%%%%%%%%
\section{\label{sec:fitting} Neural Network Training and Results}

The class of ANNs described in Section\ \ref{sec:ann} was trained using the experimental data presented earlier in Section\ \ref{sec:input_data}. The data was randomly split into training (80\%) and testing (20\%) sets. These sets were kept the same throughout this study. The training was done exclusively on the training sample. The performance of each ANN was evaluated on the test sample. 

All the features (input variables) of the model were standardized (i.e. subtracting the mean and dividing by the standard deviation) prior to training. This transformation was based solely on the training sample. This procedure ensures that the transformed features will have a mean of zero and a unit standard deviation. The exact transformation (i.e. using the mean and standard deviation obtained from the train sample) was then used on the test sample. Given the statistical nature of the train/test split the transformed test feature's mean and standard deviation might differ (slightly) from zero and unity, respectively. Though one would be tempted to use the test sample mean and standard deviation, that will bias the result, effectively defeating the purpose of having separate train/test samples. 

Given the choice of activation function for the output layer the labels (the structure function output variable) were subjected to a min/max transform  (subtracting the lowest value and dividing by the max--min range), resulting in labels in the [0,1] range. As before the procedure was based solely on the training sample labels. Once the ANN training is completed the inverse of this function needs to be applied in order to get back to the physical quantity of interest, $F_2$. The parameters associated with the scaling of both the features and the labels are saved as part of the ML model. 

The training proceeded for up to 10,000 epochs for each ANN topology. An early stopping procedure based on a minimum improvement ($10^{-6}$) in the cost function every 500 epochs was also implemented. For convenience the algorithm could start ``cold'' (i.e. random starting values for the parameters) or ``warm'' (continuing from the best set of parameters previously found for the given ANN topology). 

%As one would like to limit the bias introduced by a specific ANN architecture, an assembly of seven networks, $(5, i, i, 1)$, with $i = 40, 45, ... 70$, was used. Networks with more neurons in their hidden layers tended to overlearn, whereas going much below ten neurons/layers resulted in gross underrepresentation of the resonance region data (the fit will smooth over the resonances as there would be not enough parameters to produce the peak structures).  Given the nature of the problem deeper (more hidden layers) networks were not warranted. While individual predictions for either of the seven ANN topologies can be used a more robust estimation is the average of all ANNs. Unless otherwise specified this average represents ``the'' AI model reported here.

Several ANN networks were trained as part of this study. The number of hidden layers as well as the number of neurons per layer were varied. %{\color{red} THIS PART NEEDS REVISING!!! pending optimization studies}
The networks with less neurons/layer are more biased but also have lower variance. An increase in the number of neurons/layer results in less bias but also in a substantially larger variance. 

\begin{figure}
\begin{center}
%\hspace*{-0.75in}\includegraphics[scale=0.4]{figures/many_model_prediction_overlays1b}
\includegraphics[width=86mm]{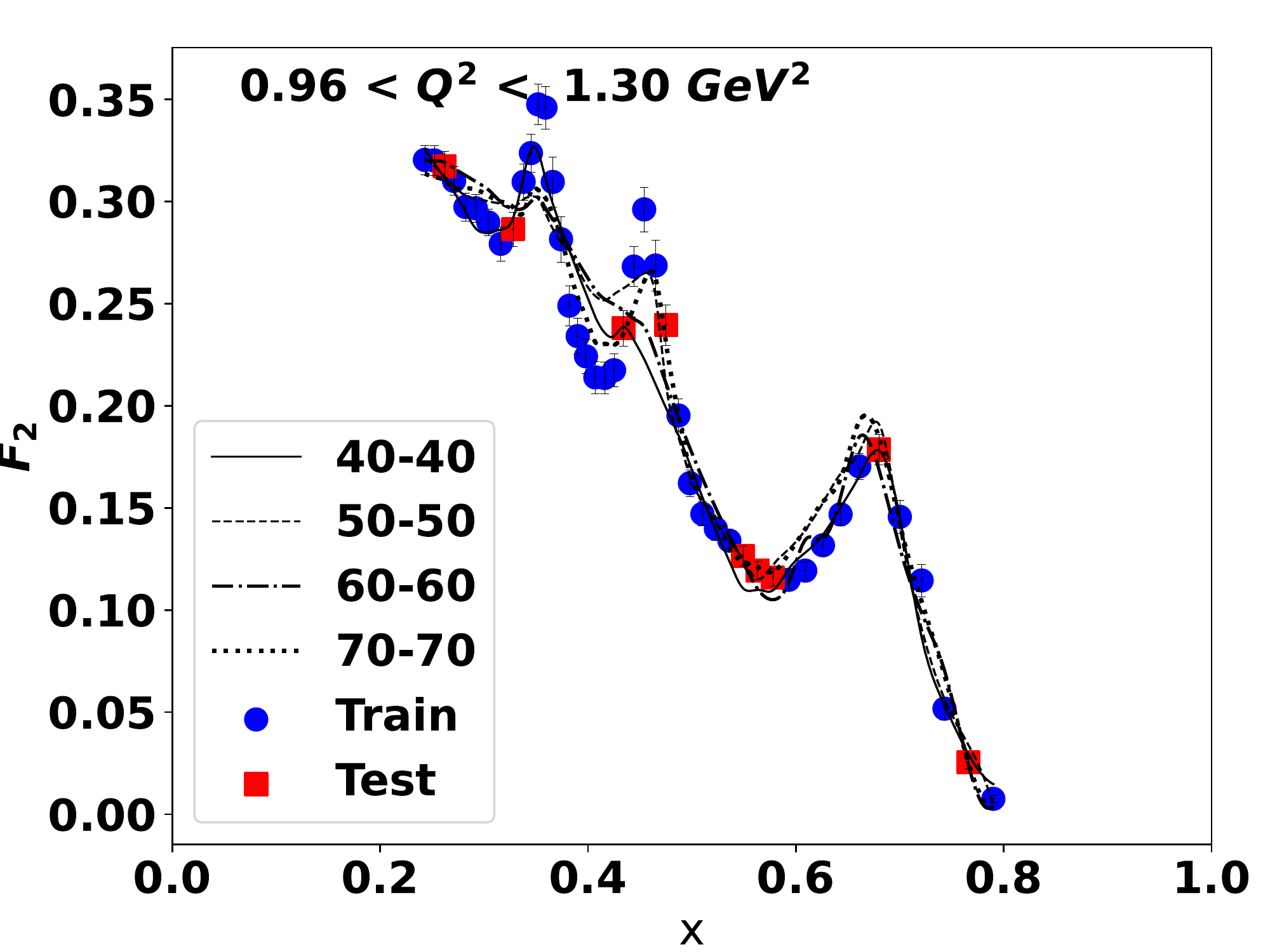}
\end{center}
  \caption{(Color online) ANN predictions for several two hidden--layer networks trained as part of this study for a sample of $F_2$ structure function data off of hydrogen in the resonance region.}
\label{fig:ann_overlay}
\end{figure}

% In this work a balance between these two extremes is achieved by averaging over several ANN predictions.
 \Cref{fig:ann_overlay} shows the ANN output for some of the two hidden--layer networks studied here for a representative set of $F_2$ (on hydrogen) data points in the resonance region. Line type (solid, dashed, etc.) and line thickness help differentiate between the various ANN topologies. The training and testing data points are shown as circles and squares, respectively. It can be seen that less complex networks have difficulties reproducing the sharper data features specific to the resonance region. For a fixed number of hidden layers, the network complexity is directly related to the number of parameters: in this study a 40-40 network has 1921 parameters while the 70-70 network has 5461 parameters. However, this study found that deeper networks\ \cite{cheap_learning} were able achieve similar or better performance with substantially less parameters. A 40-10-10-10 network has only 881 parameters and a similar mean square error to the 70-70 network.
 
In this study more than 400 different ANN topologies were trained and tested. Beside the mean square error used in the minimization procedure additional performance criteria were recorded and can be used for ``best model'' selection: the number of parameters, the mean absolute deviation (MAD) between the data and the ANN (this is a quantity less sensitive to outliers), the size of the MSE tail (defined here as the number of points for which the relative MSE is larger than 200~\%), separate MSE values for the DIS and resonance regions and also for the two targets. 

To select models with a good performance only networks with an MSE below 7\% were retained. Additionally, models were required to have less than 4~\% of the data in the MSE tail. To limit overlearning a tight cut on the difference between the training and testing MSE values was applied. Finally, only models with less than 1500 parameters were retained. This also limits the potential for overlearning while promoting faster models. Combined, these criteria selected 52 models, all of which can be used to model inclusive electron/muon scattering off of hydrogen or deuterium. 

To test the stability of the ML optimization procedure with respect to the initial choice of weights, several neural networks were repeatedly trained and their performance at the end of the training procedure was recorded. The ``cold'' training option (i.e. starting always from a random set of initial weights) was used in each case. The variation observed was of the order of 1.5~\%.

\begin{figure}
\begin{center}
\includegraphics[width=86mm]{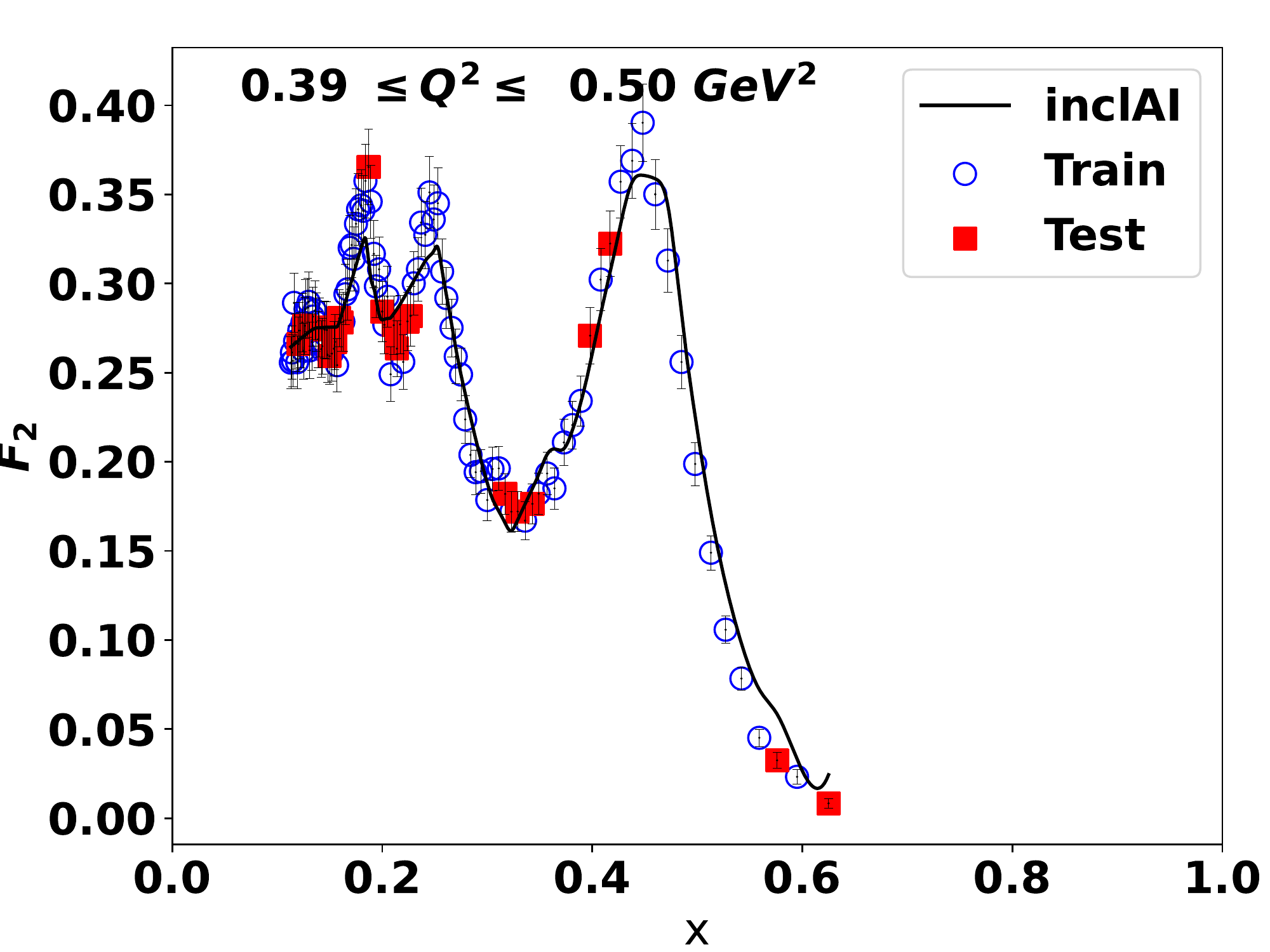}
\end{center}
  \caption{\label{fig:sample_h2_results1}(Color online) Sample inclusive AI $F_2$ results for hydrogen (I).}
\end{figure}

\begin{figure}
\begin{center}
\includegraphics[width=86mm]{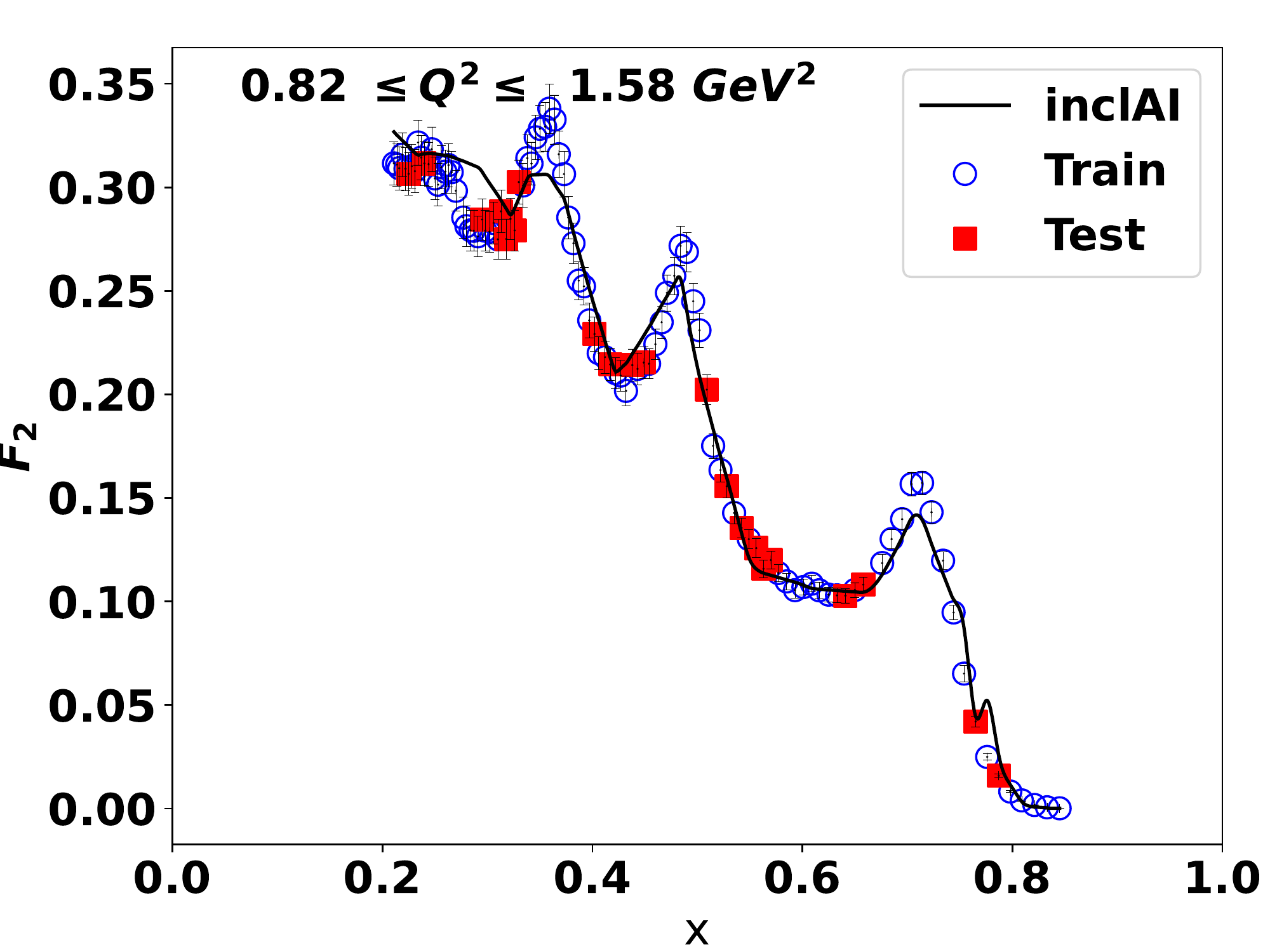}
\end{center}
  \caption{\label{fig:sample_h2_results2}(Color online)  Sample inclusive AI $F_2$ results for hydrogen (II).}
\end{figure}

\begin{figure}
\begin{center}
\includegraphics[width=86mm]{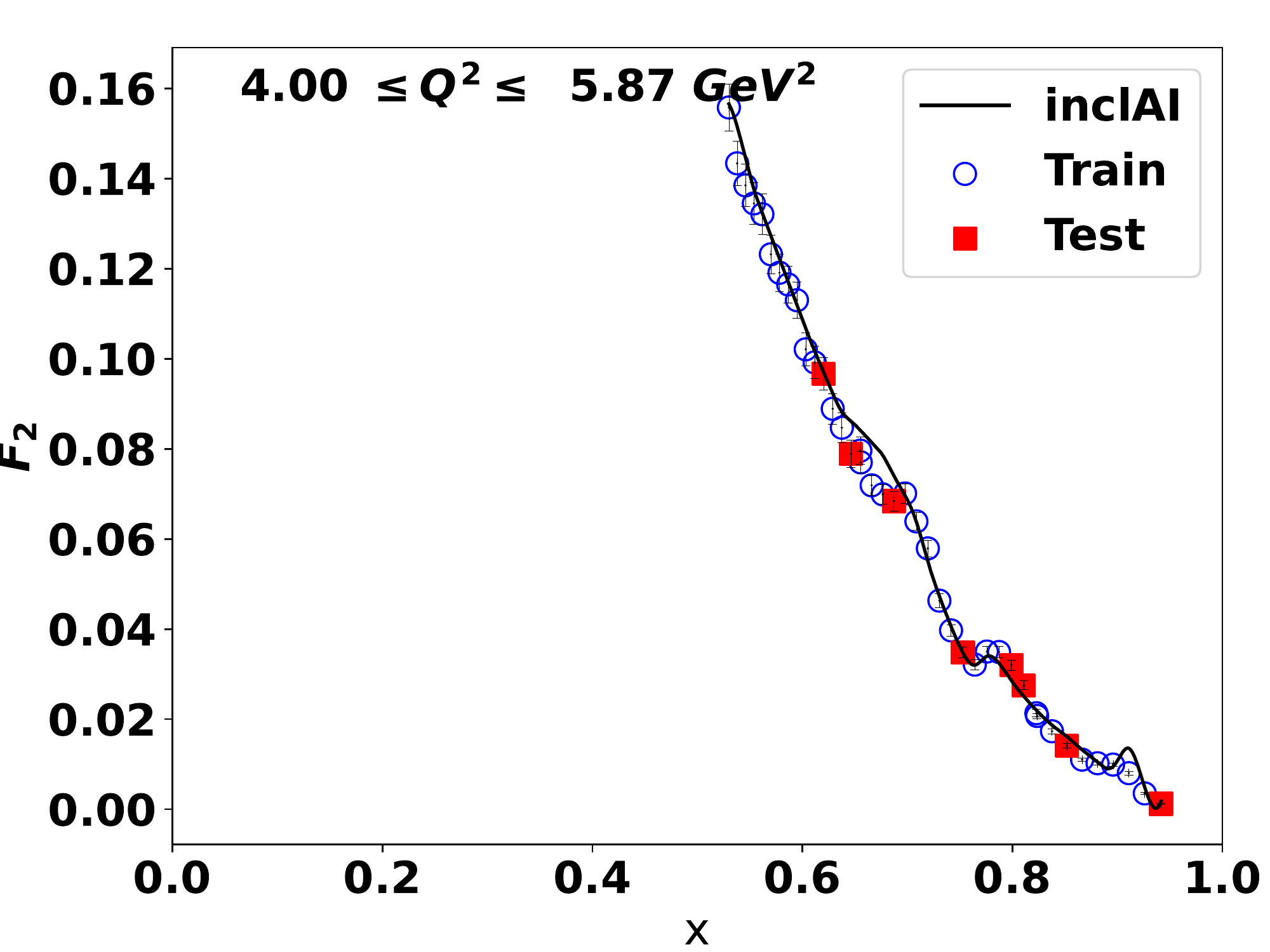}
\end{center}
  \caption{\label{fig:sample_h2_results3}(Color online)  Sample inclusive AI $F_2$ results for hydrogen (III).}
\end{figure}

\begin{figure}
\begin{center}
\includegraphics[width=86mm]{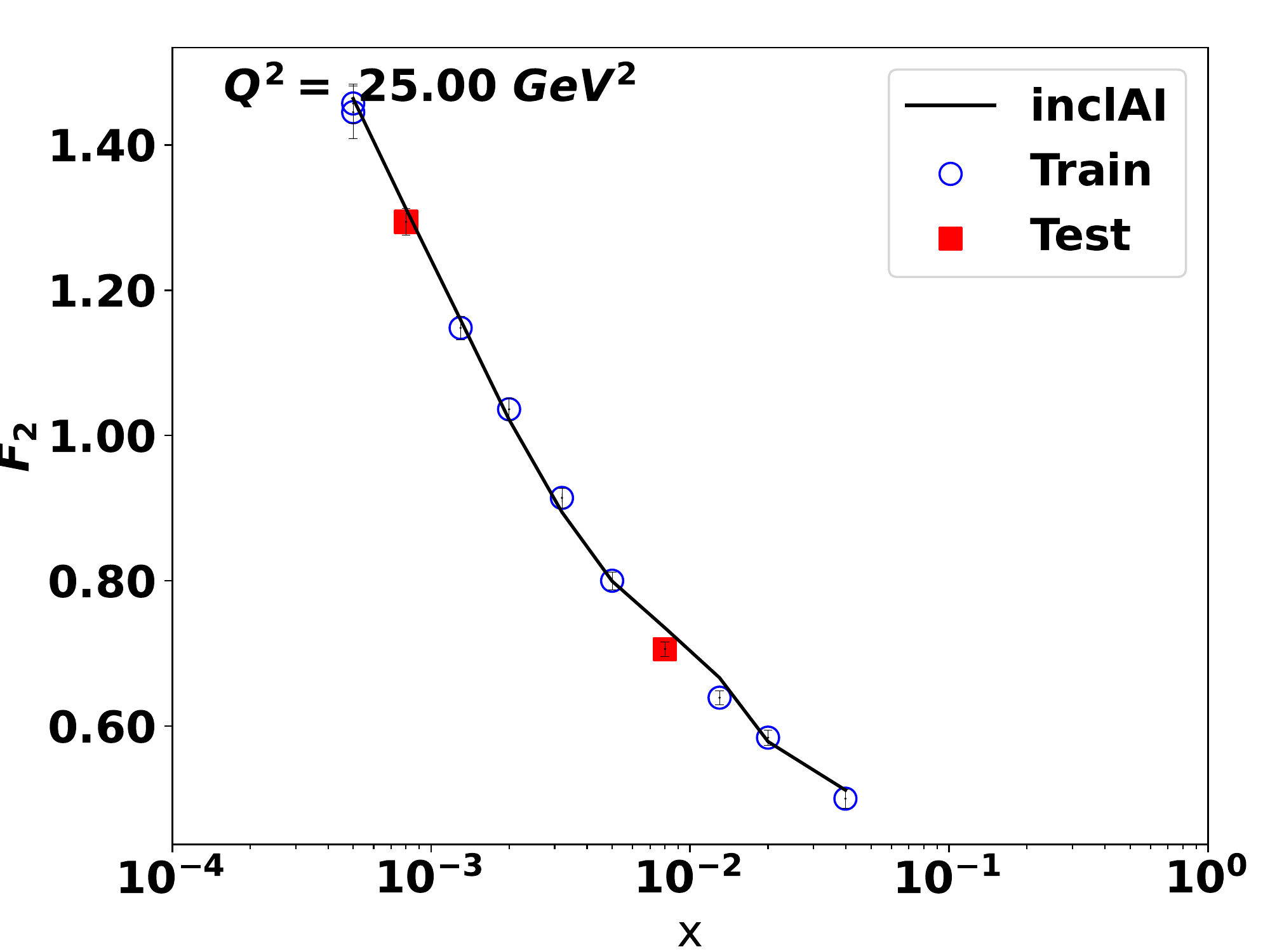}
\end{center}
  \caption{\label{fig:sample_h2_results4}(Color online)  Sample inclusive AI $F_2$ results for hydrogen (IV).}
\end{figure}

\begin{figure}
\begin{center}
\includegraphics[width=86mm]{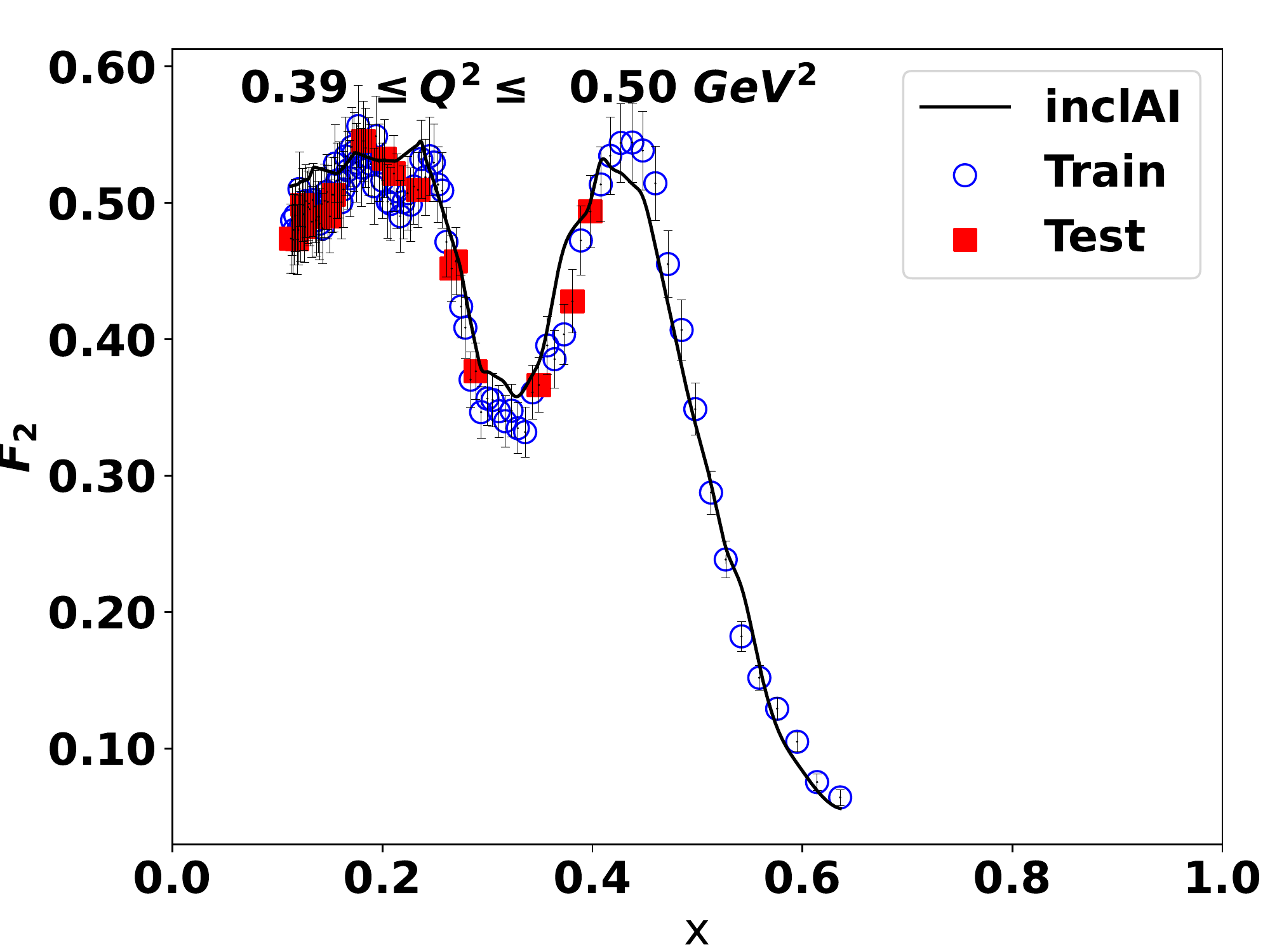}
\end{center}
  \caption{\label{fig:sample_d2_results1}(Color online)  Sample inclusive AI $F_2$ results for deuterium (I).}
\end{figure}

\begin{figure}
\begin{center}
\includegraphics[width=86mm]{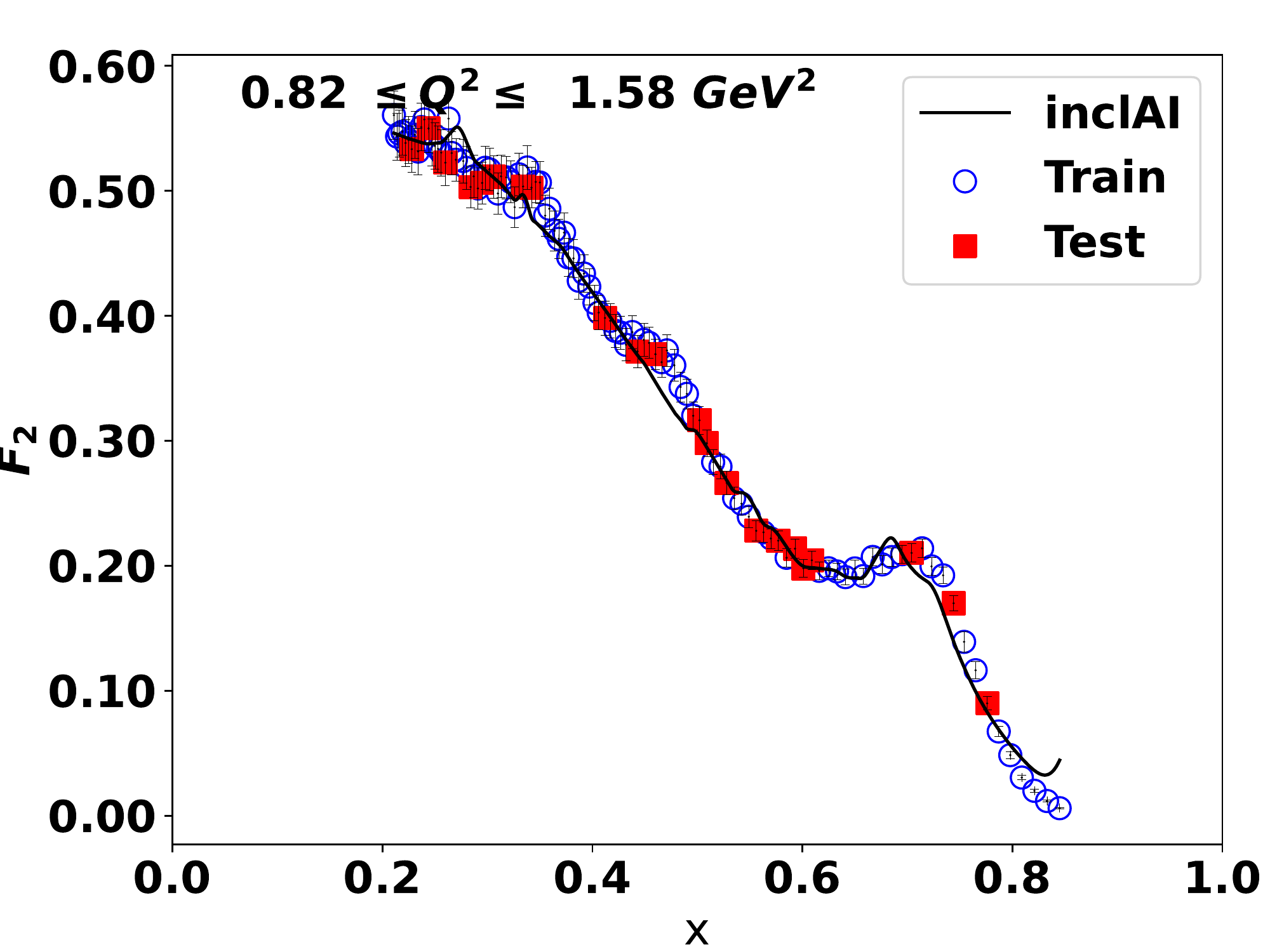}
\end{center}
  \caption{\label{fig:sample_d2_results2}(Color online)  Sample inclusive AI $F_2$ results for deuterium (II).}
\end{figure}

\begin{figure}
\begin{center}
\includegraphics[width=86mm]{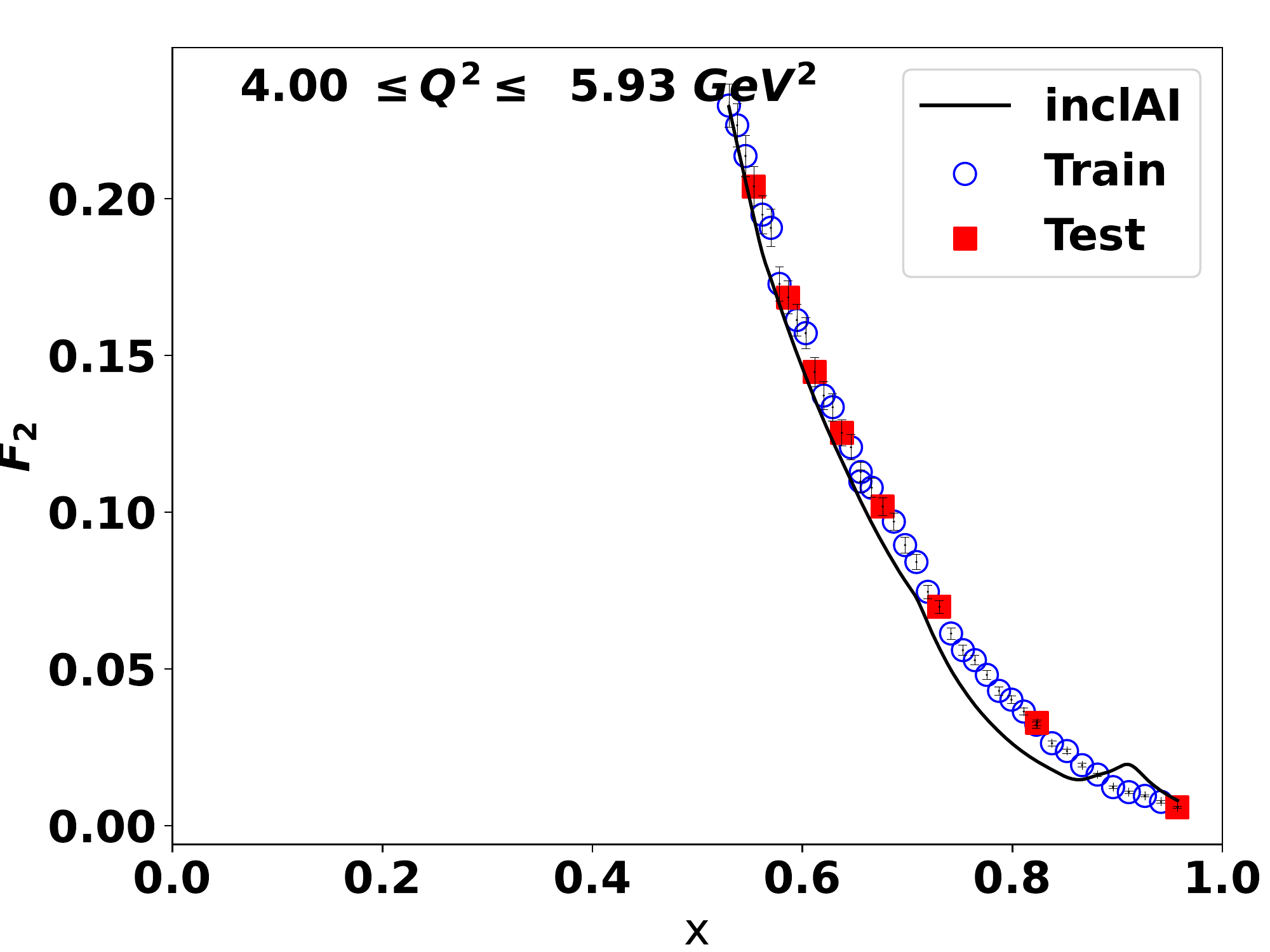}
\end{center}
  \caption{\label{fig:sample_d2_results3}(Color online)  Sample inclusive AI $F_2$ results for deuterium (III).}
\end{figure}

\begin{figure}
\begin{center}
\includegraphics[width=86mm]{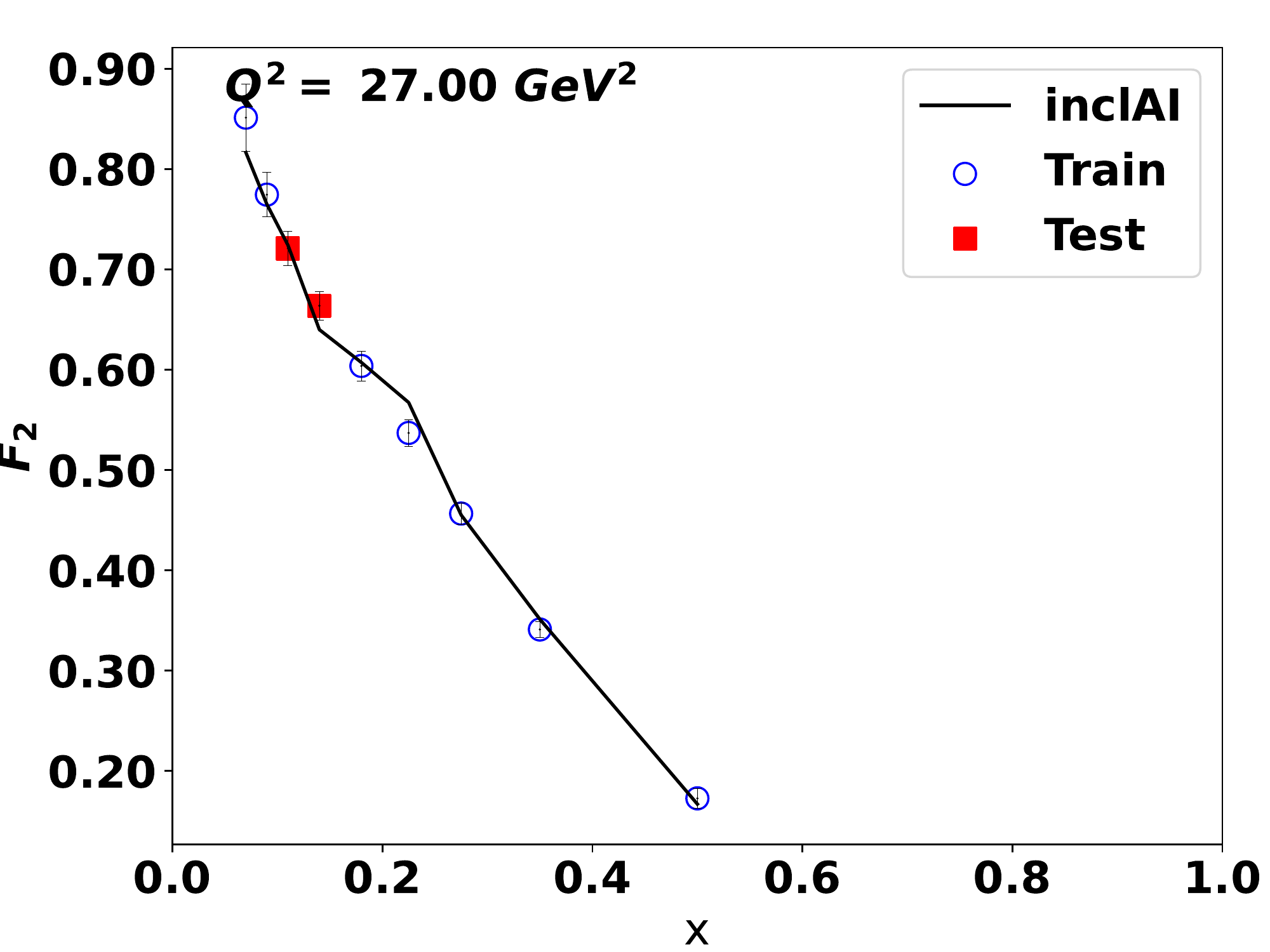}
\end{center}
  \caption{\label{fig:sample_d2_results4}(Color online)  Sample inclusive AI $F_2$ results for deuterium (IV).}
\end{figure}

\Cref{fig:sample_h2_results1,fig:sample_h2_results2,fig:sample_h2_results3,fig:sample_h2_results4} show representative hydrogen $F_2$ distributions as a function of Bjorken $x$ for various $Q^2$ ranges. For all panels the training data points are shown using circle symbols (blue in the online/color version) while the points used for testing are shown with squares (red in the online/color version). The solid line represents the AI model described in this work. Similar results for deuterium are shown in \Cref{fig:sample_d2_results1,fig:sample_d2_results2,fig:sample_d2_results3,fig:sample_d2_results4}.  For all these plots a 40-10-10-10-10-10 network was used.

\begin{figure}
\begin{center}
\includegraphics[width=86mm]{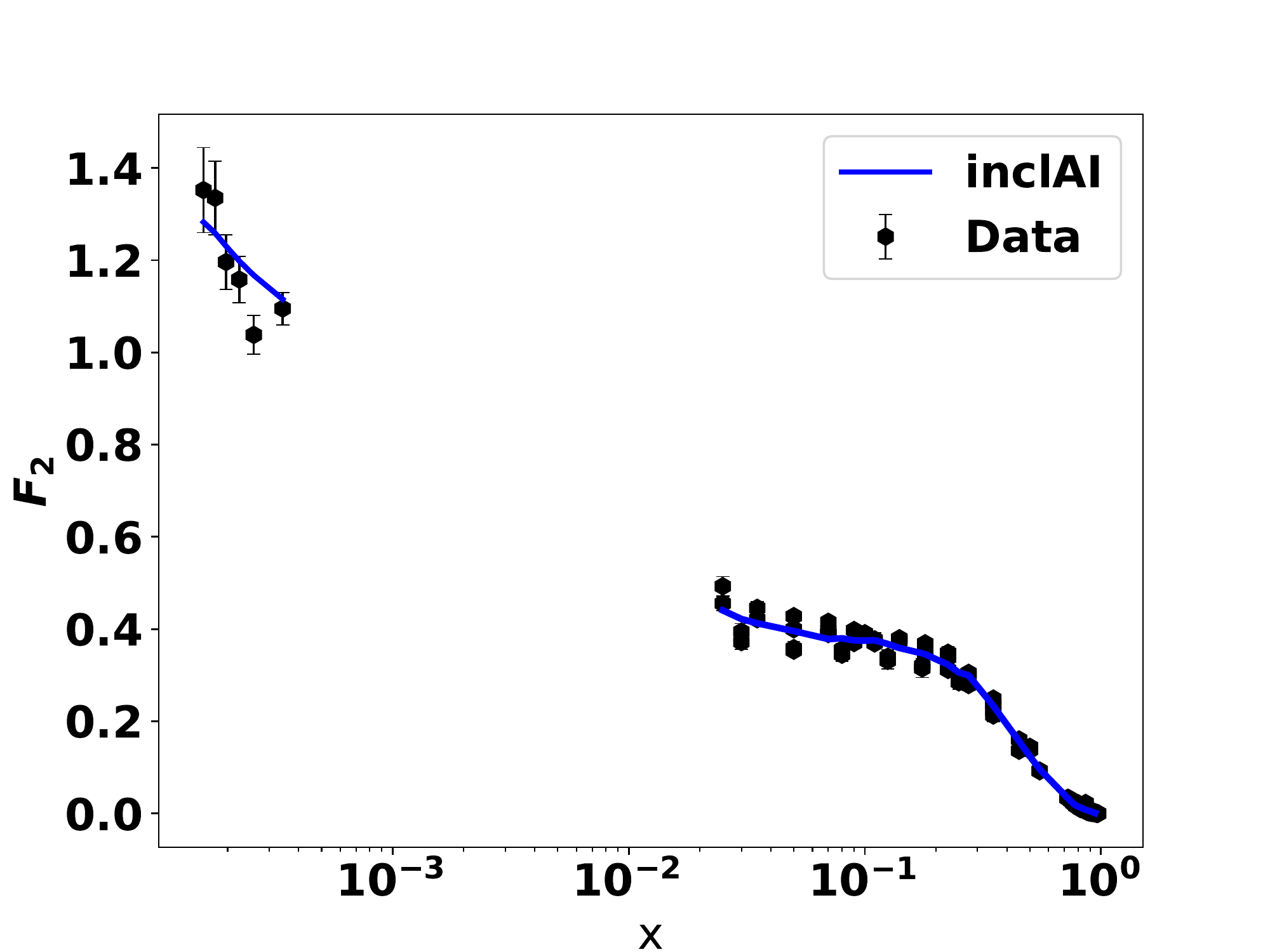}
\end{center}
  \caption{(Color online) Experimental hydrogen $F_2$ structure function (solid hexagon) compared to the inclusiveAI results (blue curve) as a function of $x$ for $7 \leq Q^2 \leq 13\; GeV^2$.}
\label{fig:f2_vs_x}
\end{figure}

One of the key strengths of the machine learning algorithm described here is its ability to model the existing experimental data over an extended kinematic range. \Cref{fig:f2_vs_x} illustrates this versatility by comparing experimental hydrogen $F_2$ structure function with the inclusiveAI model over a large $x$ range. All experimental data with $7 \leq Q^2 \leq 13\; GeV^2$  was selected. This subset of data contains results from several laboratories (SLAC, CERN, DESY, JLab).
The experimental data is shown as closed symbols, with their respective uncertainties, while the inclusiveAI calculation is represented by the shaded rectangles. It is worth reiterating that the machine learning approach uses a single set of parameters to describe all this data (as well as deuterium structure functions), whereas most theory--inspired models are more focused (and thus restrictive) on particular $x$ and $Q^2$ ranges (resonance region, DIS).

To estimate the uncertainty of the model the Monte Carlo method\ \cite{pdg, nnpdf} was used. A large (500) number of pseudo--data sets were randomly generated using the total experimental uncertainty as the standard deviation of a normal distribution centered at the nominal (published) value for each data point. The same ANN was trained on each of these pseudo--data sets, resulting in five hundred networks. These networks were used to obtain predictions for all kinematic settings used in this study. The standard deviation for each data point was then obtained and is interpreted as the model uncertainty for that point. Based on this study the overall model uncertainty due to the data errors is estimated at 4.7~\%, with slightly higher (5~\%) error in the resonance region and a lower ($\sim$3~\%) error for the DIS region.

\begin{figure}
\begin{center}
\includegraphics[width=86mm]{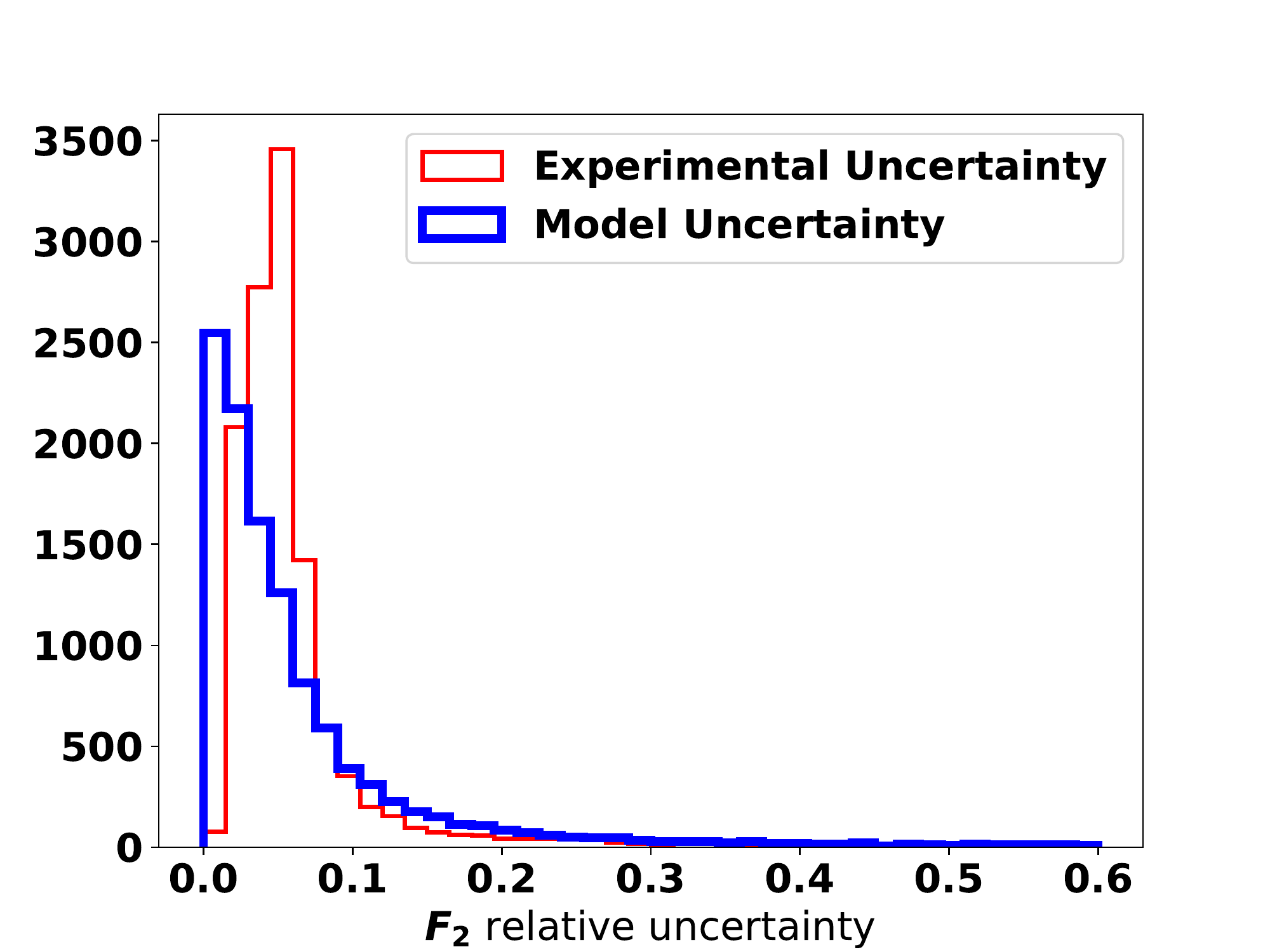}
\end{center}
  \caption{(Color online) Comparison between the total relative experimental uncertainty (thin red line) and the relative residual between the data and the inclusiveAI model (thick blue line).}
\label{fig:error_overlay}
\end{figure}

As it is the case with any machine learning project where the model is left as unbiased as possible, there is the danger of overlearning. Essentially a very complex model (very deep/wide network) ends up ``learning by heart'' the training examples but performs substantially worse on the testing set. This problem can potentially exacerbate if too many training epochs are used. To mitigate this type of problems the inclusiveAI model described here uses early stopping, a limited number of hidden layers, and possibly averaging over the predictions of several ANN topologies. \Cref{fig:error_overlay} shows a frequency distribution of the total relative experimental uncertainty for all data used in this study (thin line). The absolute value of the percent residual difference between the model prediction and the experimental data is also shown (thick line). The widths of the two distributions are similar, indicating that a) the algorithm has converged and b) the ANNs have not overlearned.

In addition, the speed of the inclusiveAI model was compared with the speed of existing models providing $F_2$ structure function predictions. As noted above, theory--inspired models make extensive use of convolutions, especially when predicting values for $A > 1 $ nuclei (deuterium in this study). The machine learning algorithm described here is a factor of 10 faster than most available models that do not rely on convolutions for their predictions and 100 times faster than models that require convolutions (for example for integrating over the Fermi motion for deuterium).

\begin{figure}
\begin{center}
\includegraphics[width=86mm]{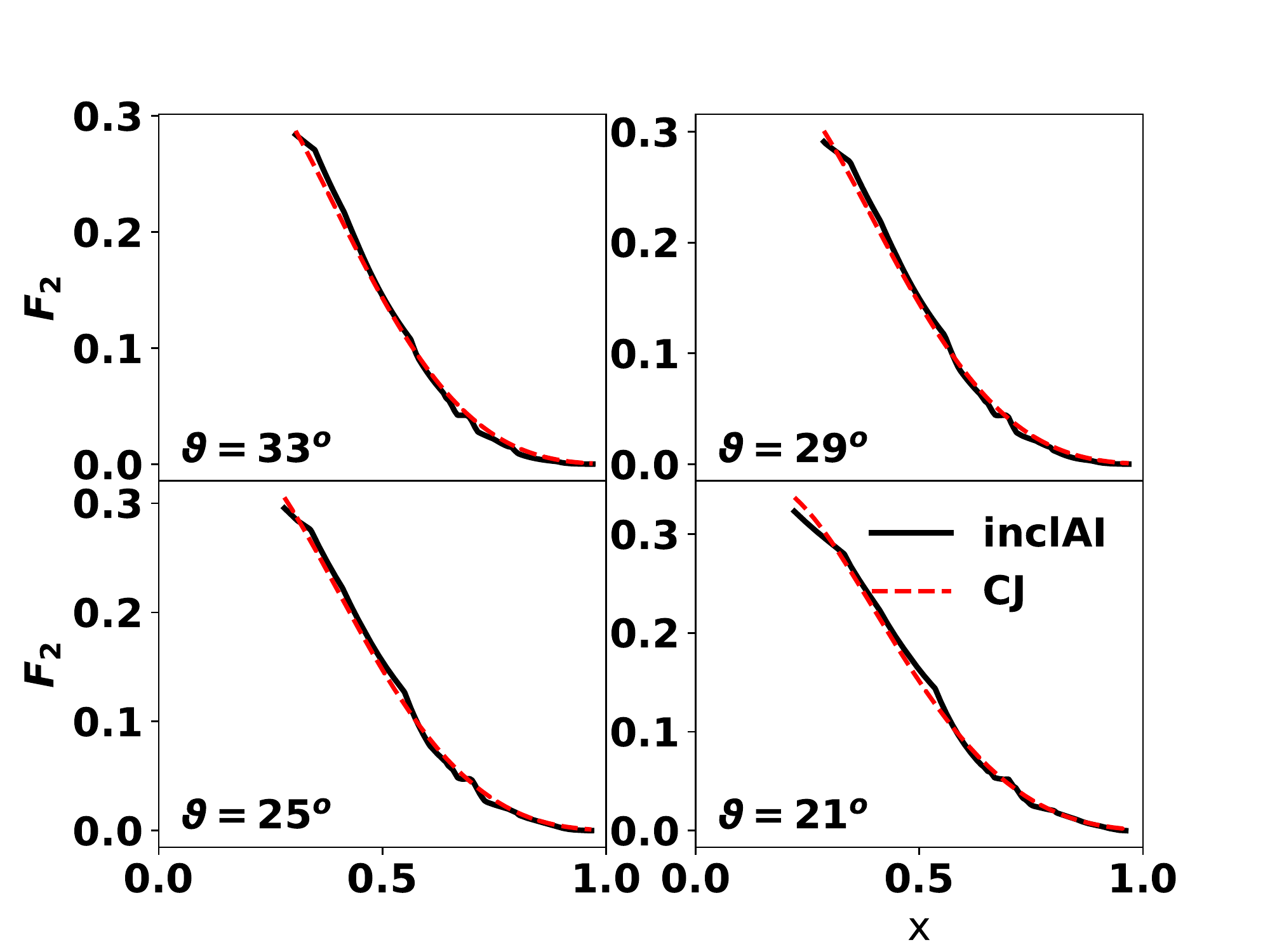}
\end{center}
  \caption{(Color online) Hydrogen $F_2$ structure function inclusiveAI predictions for some of the kinematic settings acquired by JLab Experiment E12-10-002 using a 10.6~GeV electron beam. The corresponding CJ15 predictions are shown with a dashed line.}
\label{fig:e1210002_prediction_h}
\end{figure}

\begin{figure}
\begin{center}
\includegraphics[width=86mm]{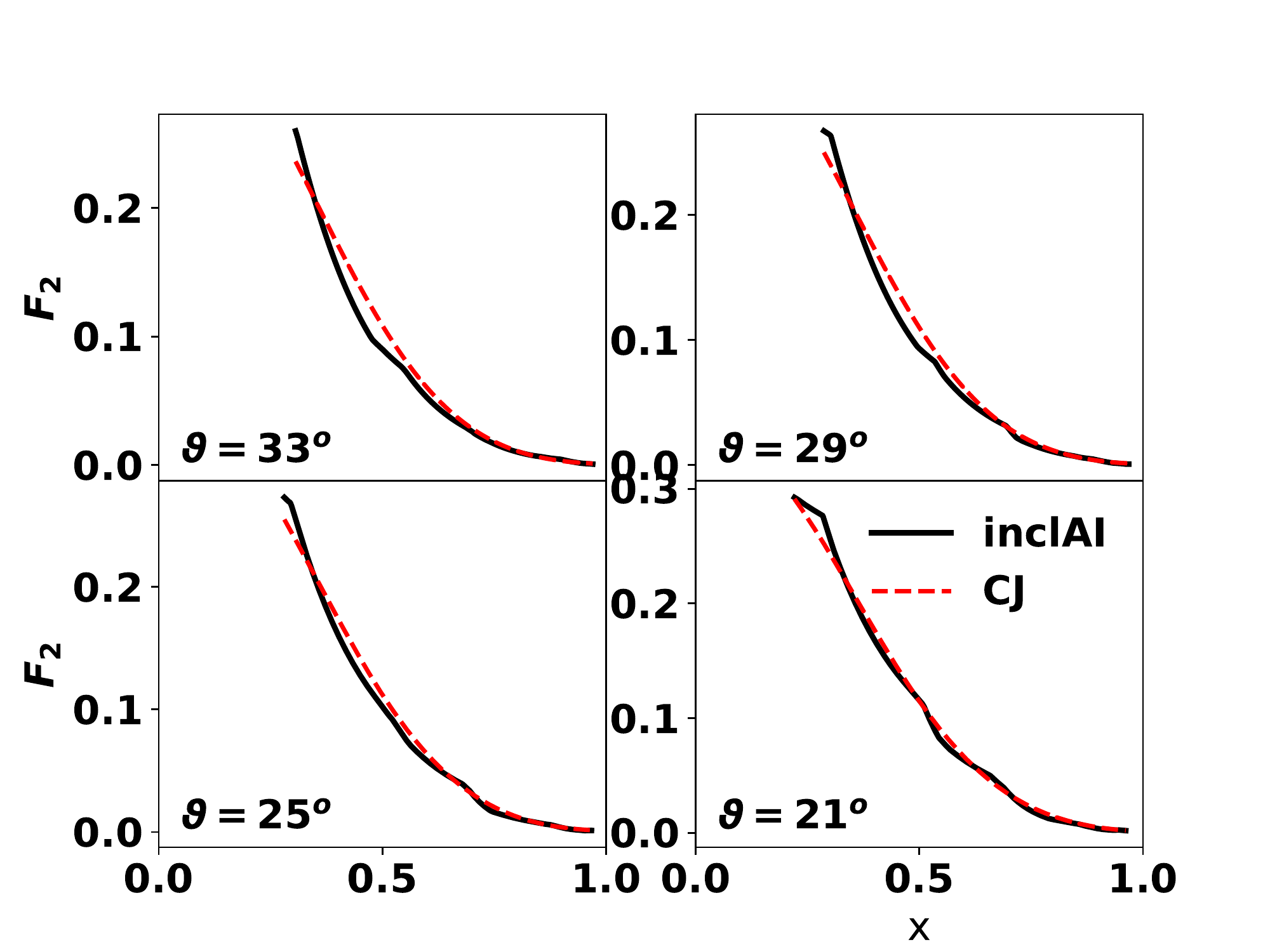}
\end{center}
  \caption{(Color online)  Deuterium $F_2$ structure function inclusiveAI predictions for some of the kinematic settings acquired by JLab Experiment E12-10-002 using a 10.6~GeV electron beam. The corresponding CJ15 predictions are shown with a dashed line.}
\label{fig:e1210002_prediction_d}
\end{figure}

Finally, to test the predictive power of inclusiveAI, the hydrogen and deuterium structure function $F_2$ was calculated for the kinematic regime probed by the JLab Experiment E12-10-002\ \cite{e1210002}, which took data in 2018 and is currently under final review before publication. \Cref{fig:e1210002_prediction_h,fig:e1210002_prediction_d} show the inclusiveAI predictions (solid line) for four spectrometer angle settings: 33, 29, 25, and 21$^o$ for hydrogen and deuterium. The data were acquired using the Super High Momentum Spectrometer (SHMS) with a 10.6~GeV electron beam. The dashed line depicts the CJ15\ \cite{cj15} calculations.

%%%%%%%%%%%%%%%
\section{\label{sec:conclusion} Conclusions.}

A machine learning model (inclusiveAI) of the $F_2$ structure function was developed. The model implements fully connected artificial neural networks with up to ten hidden layers. Its input features are $Q^2$, $W^2$, $x$, $Z$, and $A$ and its output  (label) is the $F_2$ structure function. The model was trained on the inclusive leptoproduction world data in the $x$ range from $2. \times 10^{-5}$ to the pion threshold, and in $Q^2$ from 0.055 to 800~$GeV^2$. 

With a single set of parameters the model reproduces equally well deep inelastic scattering as well as resonance data, for both hydrogen and deuterium. The mean absolute deviation between this model and the data is at 7.5~\%, comparable with the average experimental uncertainty of the global data set. Independently evaluated, the model uncertainty due to network topology and training strategy is at the 5~\% level. As the atomic and mass numbers of the target are input features, the model can be easily extended to heavier nuclei. Compared with other available structure function parameterizations, inclusiveAI is very fast, a factor of 10 faster for hydrogen predictions and typically 100 times faster for deuterium, making it an ideal candidate for event generators/Monte Carlo simulations. The Python code for defining and using (including parameters for pre-trained networks) the ML models described here is available from the authors upon request.

%%%%%%%%%%%%%%%%%
\vskip 0.25in
\noindent{\bf Acknowledgements}

This work was supported by the National Science Foundation, Grant \# 1913257.  The authors would also like to thank Mr. Thomas O'Neill for his help in reviewing the manuscript.

%%%%%%%%%%%%%%%%%
\bibliographystyle{plain}

\end{document}